\documentclass[prb,onecolumn,showpacs,preprintnumbers,amsmath,assume]{revtex4}
\usepackage{graphicx}% Include figure files
\usepackage{dcolumn}% Align table columns on decimal point
\usepackage{bm}% bold math
 \pdfoutput=1

\def\bk{ {\bf k} }

\def\bd{ {\bf d} }
\def\be{ {\bf e_z} }

\def\re{ \,{\rm Re}\, }
\def\bv{ {\bf v} }
\def\sr{ Sr$_2$RuO$_4$ }
\def\re{ \,{\rm Re}\, }

\begin{document}

\title{Electronic heat transport for a multiband superconducting gap in Sr$_2$RuO$_4$}
\author{P. Contreras}
\affiliation{Departamento de F\'{\i}sica, Universidad De Los Andes, M\'erida, 5101, Venezuela}

\date{\today}

\begin{abstract}
This paper gives a detailed numerical
study of the superconducting electronic heat transport in
the unconventional multiband superconductor Strontium Ruthenate \sr. The study
demostrates that a model with different nodal structures on
different sheets of the Fermi surface is able  to describe
quantitatively experimental heat transport data. The contribution of
the density of states DOS is given for each sheet of the Fermi surface and
the total contribution is also calculated.
Finally, a discussion of the universal character of the electronic heat transport
in unconventional superconductors and its relation to the DOS
based on the type of nodal structure of the superconducting gap in \sr is given.


\end{abstract}


\maketitle

\section{Introduction}\label{sec:intro}

It is believed that Sr$_2$RuO$_4$, a multiband superconductor with a Fermi
surface composed of three sheets, (called the $\alpha$, $\beta$ and
$\gamma$ sheets), is an unconventional superconductor with some
kind of nodes in the superconducting gap
\protect\cite{mae94,mae02}. For instance, a number of theoretical
works \protect\cite{zhi01,alt01,gra01,wu02,wys03} have predicted
the existence of line nodes on two of three sheets of the Fermi
surface ($\alpha$ and the $\beta$ sheets). While, many
authors take the $\gamma$ sheet to be nodeless, these works have
been able to provide an agreement with the specific
heat $C(T)$ \protect\cite{yasu00,deg04}, electronic heat transport experiments $\kappa(T)$
\protect\cite{mak1}, and recently with ultrasound
measurements $\alpha(T)$ as well. However, the existence of a nodeless gap for
the $\gamma$ sheet contradicts the nodal activity observed in
ultrasound measurements below T$_c$ in \sr \protect\cite{lup01}.
Firstly, the anisotropy inherent to the $\bk$ dependence of
electron-phonon interaction shows that $\gamma$ sheet
dominates ultrasound attenuation for L[100], L[110], and
T[110] sound modes. Secondly, these modes show very similar
temperature power law behavior below T$_c$; therefore
the $\gamma$ sheet should also have a similar nodal
structure.

The sound attenuation $\alpha(T)$ in \sr can distinguish the
nodal structure of the $\gamma$ sheet from that one of the
$\alpha$ and $\beta$ sheets. In contrast, electronic thermal
conductivity and specific heat have an integral effect (the three
sheets contribute to $\kappa(T)$ and $C(T)$), and it is
very difficult to discern if the order
parameter in each of the Fermi sheets has similar nodal structure.
So far, to summarize there is a considerable consensus as
to the unconventional behavior \protect\cite{mae94,mae02,ric95},
and the symmetry of the superconducting gap \protect\cite{luk98},
and probably also, to the multiband nature of the superconducting
state, nevertheless certainly there is as yet no agreement as to the nodal
structure of the superconducting gap on the different sheets of
the Fermi surface.

It has been proposed a model based on symmetry considerations
\protect\cite{wak04}, which explains the temperature behavior of
the ultrasound attenuation for the L[100], L[110], and T[110]
sound modes. According to this model, the $\gamma$ sheet shoud
have well-defined point nodes, and the $\beta$
and/or $\alpha$ bands could have also point nodes, but with an
order of magnitude smaller than for the $\gamma$ band and
resembling lines of a very small gap. The purpose of this article
is to apply, the model found in reference \protect\cite{wak04} to the study of the
electronic heat transport of \sr.

\sr has a body centered tetragonal structure
with a layered square-lattice structure similar to that of many high temperature
copper-oxide superconductors \protect\cite{mae94}. The critical temperature T$_c$ varies
strongly with non magnetic impurity concentration, T$_c$
$\approx$ 1.5 K for pure samples. The normal state displays Fermi
liquid behavior \protect\cite{mae02}. According to some authors
\protect\cite{luk98,wak05} the symmetry of the gap structure is believed to be a time
reversal broken state, with the symmetry transforming as the two
dimensional irreducible representation E$_{2u}$ of the tetragonal
point group D$_{4h}$. Additionally in reference \protect\cite{maz97} the order parameter
of \sr was proposed according to a novel mechanism due to antiferromagnetic
fluctuations.

\section{Model for the superconducting gap structure}\label{sec:model}
As I mentioned before, the gap model proposed in reference \protect\cite{wak04}
is extended here for the study of the electronic heat transport.
The assumption of a superconducting order parameter according to symmetry
considerations, where $\Delta_k$ $=$ $(\bd^i (\bk) \cdot
\bd^{i,\ast} (\bk) ) \; \Delta^i(T)$, with different $\bd^i
(\bk)$- vector order parameters for different $i$-Fermi sheets,
transforming according to the two dimensional irreducible
representation E$_{2u}$ of the tetragonal point group D$_{4h}$, it
yields the form
\begin{equation}
            \bd^i (\bk) = \be [ d_x^i (\bk) + i \: d_y^i (\bk) ];
            \label{gap}
\end{equation}
where the explicit expressions I use for $d_x^i$ and $d_y^i$
are
\begin{equation}
        \label{real_gap}
            d_x^i (\bk) =\delta^i\sin (k_x a) + \sin(\frac{k_x a}{2})
                \cos(\frac{k_y a}{2})\cos(\frac{k_z c}{2}),
\end{equation}
and
\begin{equation}
        \label{imag_gap}
            d_y^i (\bk)=  \delta^i\sin (k_y a) + \cos(\frac{k_x a}{2})
                \sin(\frac{k_y a}{2})\cos(\frac{k_z c}{2}),
\end{equation}
with d$_x^i$ and d$_y^i$ real. The factors $\delta^i$
were obtained in reference \protect\cite{wak04} by
fitting experimental data on ultrasound attenuation of reference
\protect\cite{lup01}.

\begin{figure}
\includegraphics[width = 3.0 in, height= 2.5 in]{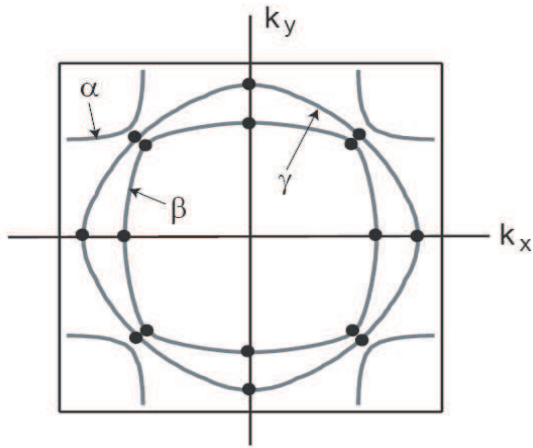}
\caption{\label{FS}The black dots show the
positions of the point nodes in the superconducting gap on the
$\beta$ and $\gamma$ Fermi surface sheets in \sr, as determined by
Eqs.~\ref{real_gap} and \ref{imag_gap}. Each solid circle
represents two nodes, at positions $\pm k_z$.}
\end{figure}

For this model, the nodal structure of the $\gamma$ band predicts
eight symmetry-related nodes for $\bk$, lying on the symmetry
equivalent $\lbrace100\rbrace$ planes (see Fig.~\ref{FS}) and
also eight symmetry-related nodes in $\lbrace110\rbrace$ planes
for the $\gamma$ sheet as well. The nodal structure of the order
parameter for the $\beta$ and/or $\alpha$ sheets yields to eight
symmetry-related nodes for $\bk$, lying on the symmetry equivalent
$\lbrace100\rbrace$ planes (see also Fig.~\ref{FS}) and also
eight symmetry-related nodes in $\lbrace110\rbrace$ planes. All
point nodes are ``accidental'' in the sense that they are not
required by symmetry, but exist only if the material parameters
$\delta^{\gamma}$ and $\delta^{\beta/\alpha}$ have values in a
certain range.

\section{Electronic heat transport}\label{thermal}

In order to calculate the electronic heat transport or thermal conductivity
for an unconventional superconductor, a standard formalism previously reported
\protect\cite{book} is extended here to account for tight
binding effects. For this purpose, the incorporation of
tight binding effects for the nodal structure and the Fermi
surface is the most important. A self-consistent calculation of
the gap function and of the energy-dependent impurity scattering
rate is beyond the scope of this article, this leads to the
approximation where the interband scattering effects are totally
neglected.

Recent induced impurity scattering measurements
\protect\cite{suz1} of the thermal conductivity for temperatures
close to 100 $mK$, have shown a remarkably universal character of the
electronic heat transport in \sr. It has been suggested
\protect\cite{suz1} that \sr is in the unitary impurity scattering
regime. In addition to this, some time ago
it was suggested the idea \protect\cite{var86}
that the transport properties of
heavy-fermion superconductors can be explained in terms of an
effective electron scattering rate which, except for the lowest
temperatures is approximate temperature independent and equal in
magnitude to that of the normal state. Such a lifetime arises in a
self-consistent treatment of impurity scattering near the strong
regime.

Thus, in the proposed not self-consistent treatment, we take the impurity
scattering superconducting quasiparticle lifetime $\tau^i_n$ to be equal to
the impurity scattering predicted for the unitary limit in the
normal state.

\begin{equation}
\frac{1}{\tau^i_n} = \frac{n_i}{\pi N^i_0 U_0}.
\end{equation}
$N^i_0$ is the density of states for the $i$-sheet at the Fermi level, and
the strength of the impurity potential $U^i_0$ $\gg$ 1.

The energy of a normal-state $i$-sheet $\epsilon^{i}_{\bf k}$ is
the same as the value previously reported \protect\cite{wak04}. The tight
binding parameters used for the $\gamma$ sheet are
$(E_0-E_F,t,t^\prime)$ = $(-0.4,-0.4,-0.12)$. These values are in
agreement with Haas-Van Alfen, and ARPES experiments
\protect\cite{ber03,maz97}. The calculation of the thermal
transport makes use of the Fermi velocity as determined from
the expression for the band structure; however, a calculation with
the unit isotropic vector of the Fermi velocity
$\hat{\bv}^i_{F,j}$ = $(\hat{j} \cdot \bk_i )$ provides the same
result.

The expression used to calculate the electronic heat transport due
to non magnetic impurity scattering in unconventional
superconductors is validated for energies $\epsilon$ $\sim$ $T$
and low impurity concentrations. A suitable equation to compare numerical calculations
with experiments is given by

\begin{equation}
    \label{at}
        \frac{\kappa_{j}(T)}{\kappa_{j}(T_c)}=  \frac{I}{T}
        \sum_i \int_0^\infty
                d\epsilon \; \epsilon \;
        \Big(-\frac{\partial f}{\partial \epsilon}\Big) \;
        A^i_{j}(\epsilon),
\end{equation}
here the constant $I$ $=$ $6/(\pi^2 T_c)$, $j$ refers to one of
the two basal directions [100] and [110] and $i$ labels the
sheets on the Fermi surface. $A^i_{j}(\epsilon)$ is given by
\begin{equation}
    \label{fat}
            A^i_{j}(\epsilon)=  \frac{\left\langle
                    \hat{\bv}^{i,2}_{F,j}(\bk) \re\sqrt{\epsilon^2-|\Delta^{i}_\bk|^2}
                        \right\rangle_{FS}}{ \sum_i \left\langle
                                \hat{\bv}^{i,2}_{F,j}(\bk)\right\rangle_{FS}}.
\end{equation}

The gap $\Delta^i_\bk$ corresponds to the expressions for
d$_x^i$ and d$_y^i$ given in Eqs.~\ref{real_gap}
and~\ref{imag_gap}. $\hat{\bv}^i_{F,j}$ are the unit Fermi
velocity vectors for each sheet. In this equation vertex
corrections have been neglected, since we have not carried out
a self-consistent evaluation of the order parameter. It is assumed a
temperature dependence of the form $\Delta^i(T)$ =
$\Delta_0^i~\sqrt{1 - (T/T_c)^3}$, which is sometimes used in the
literature \protect\cite{wu02}.

\begin{figure}
\includegraphics[width = 3.0 in, height= 2.8 in]{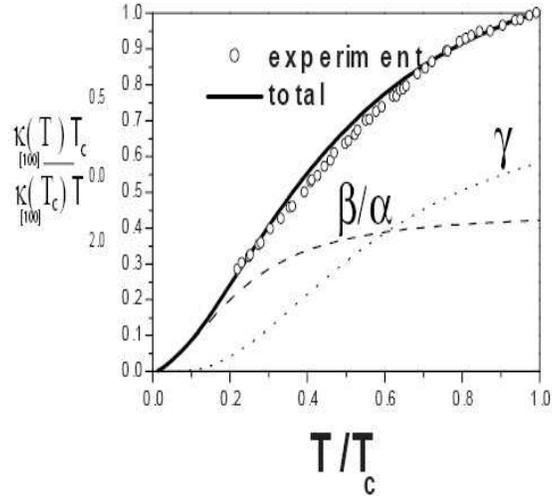}
\caption{\label{CT}Numerical fit of the in-plane thermal
conductivity components [100] normalized at T$_c$. Experimental
data is taken from \protect\cite{mak1}.}
\end{figure}

Fig.~\ref{CT} shows the numerical results for the
temperature dependence of the normalized electronic heat transport
calculated by evaluating Eq.~\ref{at}. The experimental results
are from reference \protect\cite{mak1}. The obtained fittings are:
$\Delta^\beta_0$ $\approx$ 0.09 meV, and $\Delta^\gamma_0$
$\approx$ 0.3 meV.

The numerical calculation of the electronic thermal conductivity
for the basal [100] direction is shown in Fig.~\ref{CT} The
total $\kappa$ gives an excellent agreement with the experimental data.
This experimental data is for pure samples with T$_c$ $\approx$
$1.44$ K. Contributions from each band are also calculated.
Fig. 2 shows that lines of very small point gaps on
the $\beta$ and/or $\alpha$ sheets could dominate the
behavior of the electronic thermal conductivity at low
temperatures. On the other hand, point nodes on the $\gamma$ sheet
give an insignificant contribution at low temperatures. This shows
in principle that the electronic thermal conductivity in \sr is an
integral effect where, in contrast to sound attenuation, the
contributions for all the bands are needed in order to explain the
experimental data.

It is important to notice that although the nodes in
the proposed model are both point nodes, the low temperature properties of
the nodal structure for the $\beta$ and/or $\alpha$ sheets have
some of the properties of line nodes, as there is a very low gap
along the line joining the nodes. Hence, the result match
the one provided by horizontal line nodes as is the
case of the Zhitomirsky-Rice model \protect\cite{zhi01}.
However the fit in Fig.~\ref{CT} provides in principle, a strong support for the essential features
of the order parameter symmetry model of reference \protect\cite{wak04}.

Due to the importance of anisotropy tight binding effects, I also consider worth
to calculate $\kappa$ for the [110] direction, finding that
there are no crucial differences between the calculation for any
of the two directions; even in the case when anisotropic effects are
included. This proves that the $\kappa(T)$ dependence on T (in
contrast to the sound attenuation) is in overall equally dominated
from contributions coming from all the sheets and explains the
apparent line nodes behavior of $\kappa$ where the $\beta$ and
$\alpha$ sheets give a major contribution.

\section{Density of states and universal behavior of the electronic heat transport}\label{universal}

Next, I attempt to give a qualitative analysis of the universal behavior of the
superconducting electronic thermal conductivity
according to this model. First, a numerical calculation of the density of
states DOS in the superconducting state of the $\gamma$ and
$\beta$ sheets is performed. The numerical calculation of the DOS
is made with the equation

\begin{equation}
N(\epsilon) = N^{i}_0 \; Re \Big\langle
\frac{\epsilon}{\sqrt{\epsilon^2-|\Delta^{i}_{k}|^2}} \Big
\rangle_{FS},
\end{equation}
$N^{i}_0$ is the DOS at the Fermi level for each band, and
the gap $\Delta^{i}_0$ corresponds to the one given by the
expressions for $d^{i}_x$ and $d^{i}_y$. For the purpose of
finding the energy dependence of the DOS the previous expression
is calculated at $T$ $=$ $0$.

\begin{figure}
\includegraphics[width = 3.0 in, height= 2.8 in]{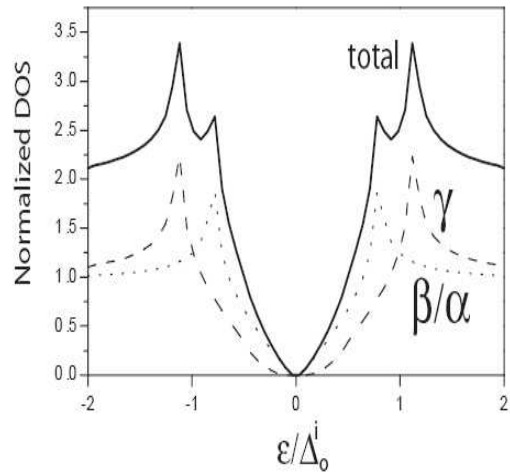}
\caption{\label{DOS} Numerical calculations of the
superconducting density of states DOS for the $\beta$ and $\gamma$
sheets according to our model.}
\end{figure}

The numerical calculation of the DOS is shown in
Fig.~\ref{DOS}. For the $\beta$ and $\gamma$ sheets
Fig.~\ref{DOS} shows that for the low energy Bogoliubov
excitations the DOS in the $\beta$ sheet behaves linearly in
energy, as it is predicted for line nodes. On the other hand, the
$\gamma$ band shows a quadratic behavior in energy, as it is
expected for a superconducting DOS with point nodes. Also from
Fig. 3 it can be noticed that the zero temperature
gap amplitude for the $\gamma$ band is larger than the one for the
$\beta$ band.

In order to elucidate the universal behavior in $\kappa$ from the
symmetry model, an intuitive approach which was found to be successful
in reference \protect\cite{zhi02} will be used below. It is noticed that a more
general theoretical analysis as the one developed for the
universal behavior of the sound attenuation in
\protect\cite{wak02} can be performed easily here. I show that it
is possible to provide the main results in the case of a
multiband superconductor. I start with a simple expression for
the expected low energy dependence of the DOS where the nodal
points (the points where the Bogoliubov quasiparticle's energy
is zero) determine the low temperature thermodynamic properties.

As it was noticed previously, the largest material parameter
$\delta{^\gamma}$ gives rise to a well defined
point-nodes topology for the order parameter in the $\gamma$ band,
therefore as shown in Fig. 3, a qualitative behavior of the DOS
yields the following expression for the density of states of the $\gamma$-sheet

\begin{equation}
\frac{N^{s,\gamma}}{N^{\gamma}_{0}} \simeq \frac{\epsilon^2}{\Delta^{\gamma,2}_0}.
\end{equation}

On the other hand, the material parameter $\delta^{\beta}$ gives
rise to a line of very small point nodes on the $\beta$ sheet and
it can be approximated  by the following expression

\begin{equation}
\frac{N^{s,\beta}}{N^{\beta}_{0}} \simeq \frac{\epsilon}{\Delta^{\beta}_0},
\end{equation}
where $\Delta^{\gamma}_0$ and $\Delta^{\beta}_0$ are the maximum
gaps for each Fermi sheet.

Close to the strong scattering regime and approaching the universal limit,
I generalize the results of reference
 ~\protect\cite{zhi02} to the case of a two gaps superconductor. In
\protect\cite{zhi02} it is supposed that the low lying Bogoliubov
excitations acquire an imaginary part $\epsilon$
$\longrightarrow$ $\epsilon$ $+$ $i \; \Gamma$. These excitations
posseses an energy independent life time $\tau$ $=$ $1/2\Gamma$ and
because of the energy uncertainty principle, the low energy
quasiparticles have a spread in energy of the order of $\Gamma$ which is called the
zero energy scattering rate ~\protect\cite{wak02}.

Therefore, close to the unitary limit the DOS for both sheets is
given by
\begin{equation}
\frac{N^{s,\gamma}}{N^{\gamma}_0} \simeq \frac{\Gamma^2_{\gamma}}{\Delta^{\gamma,2}_0},
\end{equation}
and
\begin{equation}
\frac{N^{s,\beta}}{N^{\beta}_0} \simeq \frac{\Gamma_{\beta}}{\Delta^{\beta}_0},
\end{equation}
where $\Gamma_{\gamma}$ and $\Gamma_{\beta}$ are the spreads in energy for the $\gamma$ and
$\beta$ sheets.

The low temperature specific heat $c_V$ associate with the constant DOS at low energies and temperatures
$T$ $\leq$ $\Gamma$ and for both sheets is given according to ~\protect\cite{zhi02} by

\begin{equation}
c_V = c^{\gamma}_V + c^{\beta}_V \simeq \frac{2 \pi^2}{3} \; k^2_B \; T \;
\Big[ N^{\gamma}_0 \; \frac{\Gamma^2_{\gamma}}{\Delta^2_{\gamma}}
+ N^{\beta}_0 \; \frac{\Gamma_{\beta}}{\Delta_{\beta}} \Big].
\end{equation}
where  $k_B$ is the  Boltzmann constant. This way a simple estimate of the bulk electronic thermal conductivity at low energies and for both sheets in the limit of the universal behavior (at temperatures $T$ $\leq$ $\Gamma$) is given by the generalization of equation in reference ~\protect\cite{zhi02}

\begin{equation}
\kappa = \kappa^\gamma + \kappa^\beta \simeq
c^{\gamma}_V \; v_{F,\gamma} \; l_{\gamma} + c^{\beta}_V \; v_{F,\beta} \; l_{\beta},
\end{equation}
with $l_{\gamma}$ $=$ $v_{F,\gamma}$ $\frac{\hbar}{\Gamma_{\gamma}}$ and
$l_{\beta}$ $=$ $v_{F,\beta}$ $\frac{\hbar}{\Gamma_{\beta}}$ the mean free paths for both sheets.

Finally, I found the electronic heat transport in the limit
of the universal behavior to be

\begin{equation}
\frac{\kappa}{T} \approx \frac{\pi^2 \; \hbar \; k^2_B}{3} \; \Big[v^{2}_{F,\beta}
\frac{N^{\beta}_0}{\Delta^{\beta}_0} + v^{2}_{F,\gamma}
(\frac{N^{\gamma}_0 \; \Gamma_{\gamma}}{\Delta^{\gamma,2}_0})\Big].
\end{equation}

A numerical evaluation of the previous equation can be performed by using the values for
the Fermi velocities $v_{F,\beta}$ $=$ 95.9 Km/seg and $v_{F,\gamma}$ $=$ 59.5 Km/seg according to ~\protect\cite{mae02}, the values for $\Delta^{\gamma}_0$
and $\Delta^{\beta}_0$ given previously in this paper,
and the values of the DOS at the Fermi level
of reference ~\protect\cite{mae02}. For the spread in energy in the unitary limit,
I take a very simple approximation $\Gamma_{\gamma}$ $\sim$
$\Delta^{\gamma}_0$ following the appendix of reference ~\protect\cite{wak02}.

Therefore, I find the bulk thermal conductivity for this two sheets model
approximately equal to $\kappa_{theoretical}/T$ $\approx$ 0.8 $W/K^2 m$
and which is lower than the experimental value found
in reference ~\protect\cite{suz1} of $\kappa_{experimental}/T$ $\approx$ 1.7 $W/K^2 m$.

\section{Final remarks}\label{conclu}

It is found that with two different gap structures
characterized in the $\gamma$, $\beta$ and/or $\alpha$ sheets by
point nodes of different magnitude, the calculation of the
temperature behavior of the superconducting electronic
heat transport $\kappa(T)$ leads
to an excellent agreement with the existent experimental data
\protect\cite{mak2}. These results also show that in contrast to
the ultrasound attenuation $\alpha(T)$, which is able to identify different
nodal structures in different bands, there is not relevant
anisotropy for the quantities $\kappa_{[100]}$ and
$\kappa_{[110]}$. This makes difficult to identify the nodal
structure in \sr from electronic thermal conductivity data.
Finally, the numerical calculation of the DOS for each sheet,
allows us to qualitatively estimate the value of the universal limit
for the electronic heat transport to be lower than
the value observed experimentally in reference \protect\cite{suz1}.

\section*{Acknowledgments}

I thank Dr. M. Tanatar  for stimulating discussions and for
providing the experimental data in Fig.~\ref{CT}. I also
acknowledge discussions with Prof. Michael Walker from the University of Toronto,
Prof. Jos\'e Rodriguez at SUPERCOM from la Universidad de Carabobo,
and Prof. Rodrigo Casanova. This research was supported by the Grant CDCHTA
number C-1479-07-05-AA.

%===================================================================


\begin{thebibliography}{99}

\bibitem{mae94} Y. Maeno, H. Hashimoto, K. Yoshida, S. Nishizaki, T.
Fujita, J. G. Bednorz, and F. Lichtenberg, Nature {\bf 372} (1994) 532.

\bibitem{mae02} A. Mackenzie, and Y. Maeno, Rev. of Modern Physics
{\bf 75} (2003) 657.

\bibitem{zhi01} M. Zhitomirsky and T. M. Rice, Phys. Rev. Lett. {\bf 87} (2001)
057001.

\bibitem{alt01} D. Agterberg, T. M. Rice, and M. Sigrist, Phys. Rev. Lett. {\bf 78} (1997)
3374.

\bibitem{gra01} M. J. Graf and A. V. Balatsky, Phys. Rev. B {\bf 62} (2000)
9697.

\bibitem{wu02} W. C. Wu and R. Joynt, Phys. Rev. B {\bf 65} (2002)
104502-1.

\bibitem{wys03} K. I. Wysoki\'{n}ski, G. Litak, J. F. Annett,
and B. L. Gy\"{o}rffy, Phys. Stat. Sol (b) {\bf 236} (2003) 325.

\bibitem{yasu00} Y. Hasegawa, K. Machida, and M. Ozaki,  J. Phys. Soc. Jpn.
{\bf 69} (2000) 336.

\bibitem{deg04} K. Deguchi, Z. Q. Mao, H. Yaguchi, and Y. Maeno, Phys. Rev.
Lett. {\bf 92} (2000) 047002-1-4.

\bibitem{mak1} M. Tanatar, S. Nagai, Z. Q. Mao, Y. Maeno, and T. Ishiguro,
Phys. Rev. B {\bf 63} (2001) 064505.

\bibitem{lup01} C. Lupien, W. A. MacFarlane, C. Proust, L.
Taillefer, Z.~Q.~Mao, and Y. Maeno, Phys. Rev. Lett. {\bf 86} (2001)
5986.

\bibitem{wak04} P. Contreras, M. Walker, and K. Samokhin,
Phys. Rev. B {\bf 70} (2004) 184528.

\bibitem{ric95} T. M. Rice and M. Sigrist, J. Phys. Condens. Matter
{\bf 7} (1995) L643.

\bibitem{luk98} G. M. Luke, Y. Fudamoto, K. M. Kojima, M. I. Larkin,
J. Merrin, B. Nachumi, Y. J. Uemura, Y. Maeno, Z. Q. Mao, H.
Nakamura, and M. Sigrist, Nature {\bf 394}, (1998) 558.

\bibitem{wak05} M. Walker and P. Contreras,
Phys. Rev. B {\bf 66} (2002) 214508.

\bibitem{ber03} C. Bergemann, A. P. Mackenzie, S. R. Julian,
D. Forsythe, and E. Omichi, Adv. Phys.{\bf 53} (2003) 639.

\bibitem{maz97} I.I. Mazin and D. J. Singh, Phys. Rev. Lett. {\bf 79} (1997)
 733.

\bibitem{wak02} M. Walker, M. Smith, and K. Samokhin,
Phys. Rev. B {\bf 65} (2002) 014517.

\bibitem{book} V. P. Mineev and K. V. Samokhin, {\em Introduction to
Unconventional Superconductivity} (Gordon and Breach, Amsterdam,
1999).

\bibitem{zhi02} M. Zhitomirsky and M. Walker, Phys. Rev. B {\bf 57} (1998)
8560.

\bibitem{var86} S. Schmitt-Rink, K. Miyake, and C. M. Varma, Phys.
Rev. Lett. {\bf 57} (1986) 2575.

\bibitem{mak2} M. Tanatar, M. Suzuki, S. Nagai, Z. Q. Mao, Y. Maeno, and T. Ishiguro,
Phys. Rev. Lett. {\bf 86} (2001) 2649.

\bibitem{suz1} M. Suzuki, M. Tanatar, N. Kikugawa, Z. Q. Mao, Y. Maeno, and T. Ishiguro,
Phys. Rev. Lett. {\bf 88} (2002) 227004.
\end{thebibliography}
\end{document}